\documentclass{emulateapj}

\newcommand{\mbh}{$M_{\rm BH}$}
\newcommand{\sigs}{$\sigma_{\ast}$}
\newcommand{\msig}{\mbh-\sigs}

\shorttitle{Velocity Dispersion of a Luminous Quasar Host}

\begin{document}

\title{First Stellar Velocity Dispersion Measurement of a Luminous Quasar Host with Gemini North Laser Guide Star Adaptive Optics}

\author{Linda~C.~Watson\altaffilmark{1},
  Paul~Martini\altaffilmark{1,2}, Kalliopi~M.~Dasyra\altaffilmark{3},
  Misty~C.~Bentz\altaffilmark{4}, Laura~Ferrarese\altaffilmark{5},
  Bradley~M.~Peterson\altaffilmark{1,2},
  Richard~W.~Pogge\altaffilmark{1}, Linda~J.~Tacconi\altaffilmark{6}
}

\altaffiltext{1}{Department of Astronomy, The Ohio State University,
  140 West 18th Avenue, Columbus, OH 43210; 
watson@astronomy.ohio-state.edu}
\altaffiltext{2}{Center for Cosmology and AstroParticle Physics, The
  Ohio State University, 191 West Woodruff Avenue, Columbus, OH 43210}
\altaffiltext{3}{Spitzer Science Center, California Institute of
  Technology, Mail Code 2206, 1200 East California Boulevard,
  Pasadena, CA 91125}
\altaffiltext{4}{Department of Physics and Astronomy, University of
  California at Irvine, 4129 Frederick Reines Hall, Irvine, CA 92697}
\altaffiltext{5}{Herzberg Institute of Astrophysics, National Research
  Council of Canada, 5071 West Saanich Road, Victoria, BC V9E 2E7,
  Canada}
\altaffiltext{6}{Max Planck Institut f${\rm \ddot{u}}$r extraterrestrische
  Physik, D-85741 Garching, Germany}

\begin{abstract}
We present the first use of the Gemini North laser guide star adaptive
optics (LGS AO) system and an integral field unit (IFU) to measure the 
stellar velocity dispersion of the host of a luminous quasar.  The
quasar PG1426+015 ($z=0.086$) was observed with the Near-Infrared
Integral Field Spectrometer (NIFS) on the 8m Gemini North telescope in
the H-band as part of the Science Verification phase of the new ALTAIR
LGS AO system.  The NIFS IFU and LGS AO are well suited for host
studies of luminous quasars because one can achieve a large ratio of
host to quasar light.  We have measured the stellar velocity
dispersion of PG1426+015 from  $0.1''$ to $1''$ (0.16~kpc to 
1.6~kpc) to be $217\pm15 \, {\rm km \, s^{-1}}$ based on high
signal-to-noise ratio measurements of Si I, Mg I, and several CO
bandheads. This new measurement is a factor of four more precise than
a previous measurement obtained with long-slit spectroscopy and good,
natural seeing, yet was obtained with a shorter net integration
time. We find that PG1426+015 has a velocity dispersion that places it
significantly above the \msig\ relation of quiescent galaxies and
lower-luminosity active galactic nuclei with black hole masses
estimated from reverberation mapping.  We discuss several possible
explanations for this discrepancy that could be addressed with similar
observations of a larger sample of luminous quasars.
\end{abstract}


\section{Introduction}
\label{sec:intro}
The link between central supermassive black holes and the properties
of their host galaxies has been firmly established in both quiescent
galaxies and active galactic nuclei \citep[AGNs;][]{kormendy95, fm00,
g00a, g00b, f01, graham01, marconi03, haring04}.  Arguably, the
tightest correlation is the \msig\ relation, which relates the mass of
the central black hole (\mbh) and the stellar velocity dispersion of
the host spheroid ($\sigma_{\ast}$).  In active galaxies, the stellar
velocity dispersion is typically measured using the \ion{Ca}{2}
triplet stellar absorption features at rest wavelengths of 8498, 8542,
and 8662\AA. The most direct and broadly applicable method for
measuring the central black hole mass in active galaxies is through
reverberation mapping, where the time delay between continuum and
emission line variations is used as a measure of the radius of the
broad-line region ($R_{\rm BLR}$) and the width of the H$\beta$
emission line ($\Delta V$) is used as a measure of the gas velocity
within the BLR \citep{blandford82, peterson93}.  Then one uses the
virial equation to calculate the mass:
\begin{equation}
\label{eqn:virial}
M_{\rm BH}=f \frac{R_{\rm BLR} (\Delta V)^{2}}{G}.
\end{equation}
The scale factor $f$ accounts for the unknown geometry, kinematics, and
inclination of the BLR.  \citet[][hereafter O04]{onken04} estimated
$\langle f \rangle$ statistically for the AGN population under the
assumption that the zero-point of the \msig\ relation is the same for
active and inactive galaxies. They derive a value of $f=5.5 \pm 1.7$
from a sample of 14 active galactic nuclei (AGNs) with masses
determined by reverberation mapping and stellar velocity
dispersions determined using the \ion{Ca}{2} triplet.  Using
single-epoch spectroscopy, this scale factor is applied in black hole
mass estimates for large samples of AGNs out to high redshift (see
McGill et al. 2008 for a recent summary).

The calculation of $\langle f \rangle$ has unfortunately been limited
to low-luminosity AGNs for two reasons: dilution of stellar features
by the AGN continuum and the relative scarcity of higher-luminosity
AGNs.  In particular, observations of higher-luminosity objects
necessitate observations of higher-redshift objects, yet telluric
absorption mostly limits the utility of the \ion{Ca}{2} triplet to $z
\le 0.06$.  Both of these problems can be somewhat circumvented with
stellar velocity dispersion observations using CO bandheads in the
near infrared (NIR) H-band.  This wavelength range corresponds to the
maximum ratio of the stellar emission from the host galaxy to the
quasar continuum \citep{wright94,elvis94}. Additionally, current
adaptive optics (AO) systems function well in the NIR. This
combination facilitates velocity dispersion measurements of luminous
quasar hosts, which are essential to determine if these objects fall
on the \msig\ relation and/or are characterized by the same $\langle
f \rangle$ as lower-luminosity AGNs. 

A suitable sample of higher-luminosity quasars for such a study is the
reverberation-mapped sample discussed by \citet{peterson04}.  The 16
PG quasars studied in this work are on average 40 times more luminous,
with black holes 10 times more massive, than the AGNs in the O04
sample (the O04 and PG quasar samples have average luminosities of
${\rm log} (\lambda L_{\lambda}(5100 {\rm \AA}) / {\rm erg~s^{-1}}) =
43.6$ and $45.2$, respectively, and average black hole masses of
${\rm log} (M_{\rm BH}/M_{\odot}) = 7.75$ and 8.65, respectively).
Work has already begun to measure velocity dispersions in this
luminosity regime. \citet[][hereafter D07]{dasyra07} investigated the
hypothesized evolutionary link between PG quasars and ultra-luminous 
infrared galaxies (ULIRGs) with H-band CO bandhead measurements of 12
PG quasars. These data were obtained using the ISAAC long-slit
spectrometer \citep{moorwood98} on the 8m Antu unit of the Very Large
Telescope under good natural seeing conditions.  The H-band CO 
absorption features were detected in most of the quasars, including
four reverberation-mapped quasars.  However, the observations were
still of faint hosts with significant nuclear emission contamination.
This resulted in rather noisy host-galaxy spectra for some objects and
velocity dispersions with uncertainties ranging from $18$ to $67 \,
{\rm km \, s^{-1}}$ ($12 - 36 \%$ error).

Over the next few years, integral field units (IFUs) combined with
increasingly reliable AO systems should lead to substantial
improvements in studies of the hosts of luminous quasars. The primary
advantage of using an IFU is that more of the light in the region of
interest (for example, we probe the central $3\arcsec\times3\arcsec$)
is dispersed rather than just the light within a normal single slit
(e.g., $0.6\arcsec\times120\arcsec$ in D07). Consequently, more
host-galaxy light from near the galaxy's center is gathered in a
single exposure. AO is a further aid because the contamination by the
intrinsically point-like quasar can be confined to the central few
pixels of the image, thus minimizing the quasar dilution of the
stellar absorption features.  As a demonstration  of this approach, we
present observations of PG1426+015 obtained with the Near-Infrared
Integral Field Spectrometer (NIFS) during Science Verification Time
for the new ALTAIR Laser Guide Star (LGS) AO system at Gemini North.
We specifically chose PG1426+015 because it had a relatively uncertain
velocity dispersion measurement in the earlier study of D07, yet one
which placed it suggestively above the \msig\ relation (along with two
of the three other luminous reverberation-mapped quasars).

In the next section we describe our observations, data reduction, and
velocity dispersion measurement technique.  We discuss our results and
their implications in Section~\ref{sec:disc} and summarize our
findings in Section~\ref{sec:conc}.  Throughout this paper, we adopt
$H_{0}=70\, {\rm km\, s^{-1}}$, $\Omega_{\rm m}=0.3$, and $\Omega_{\rm
total}=1$.

\section{Observations, Data Reduction, and Data Analysis}
\label{sec:obs}

The observations of PG1426+015 and the velocity template HIP 75799
(K0III) were carried out at Gemini North on 2007 February 6-7, and April
29 using NIFS \citep{mcgregor02}.  Three additional velocity
templates -- HD 84769 (K5III), V* BU CVn (M1III), and BD+23 1138
(M5Ia) -- were observed on 2008 February 12, 14, and 15.  The 
image slicer of NIFS divides the 3$\arcsec\times3\arcsec$ field of
view into 29 spectroscopic slices.  NIFS was fed by ALTAIR, the AO
system at Gemini North, using both a natural and laser guide star
\citep{herriot00,boccas06}. The quasar was used as the natural guide
star for tip/tilt corrections.  The average AO-assisted seeing
during our observations was $0.14\arcsec$ in the H band.  The average
uncorrected seeing, as measured by ALTAIR at the end of each exposure,
was $0.5\arcsec$.  Typical Strehl ratios for H-band observations
using the Gemini LGS AO system are about 10\%.

Our observations were obtained with the H grating and the JH filter.
The spectral resolution is $\lambda/\delta \lambda=5710$ (FWHM velocity
resolution $\sim50 \, {\rm km \, s^{-1}}$) and there are 1.9 pixels per
resolution element. The H-band wavelength coverage from $1.485$ to
$1.80 \mu{\rm m}$ contains many strong atomic and molecular 
features, including \ion{Mg}{1} 1.488$\mu{\rm m}$, \ion{Mg}{1}
1.503$\mu{\rm m}$, CO(3-0) 1.558$\mu{\rm m}$, CO(4-1)
1.578$\mu{\rm m}$, \ion{Si}{1} 1.589$\mu{\rm m}$, CO(5-2) 1.598$\mu{\rm
m}$, and CO(6-3) 1.619$\mu{\rm m}$. 

We obtained a total of 2.08 hours integration on PG1426+015 with
sequences of 300 second exposures interspersed with regular sky
observations.  For telluric correction, an A0V star was observed at
least once for every hour of quasar or velocity template observations,
including sky frames. We also obtained the recommended calibration
frames.

The data were processed with Gemini IRAF packages and the
recommended reduction steps given on the NIFS
website\footnote{http://www.gemini.edu/sciops/instruments/nifs/}.  We 
only deviated from this prescription to remove the Brackett absorption
features in the A0V stellar spectra.  For this procedure, we used the {\it
xtellcor\_general} task written by \citet{vacca03}.

Our final host-galaxy spectrum is the difference between a 1$\arcsec$
(1.6~kpc) and 0.1$\arcsec$ radius extraction. We empirically chose the
outer radius because there was little signal beyond this point and sky
subtraction residuals became more evident when it was increased.  We
chose the inner radius because it included as much host-galaxy light
as possible, yet minimized the quasar contribution.  The LGS system
allowed our inner radius to be substantially smaller than would have
been possible under natural seeing conditions.

We measured the stellar velocity dispersion using the penalized
pixel fitting (pPXF) method of \citet{cappellari04}.  As is typical in
velocity dispersion measurements, this method assumes that the host-galaxy
spectrum is represented by a convolution of a stellar template
spectrum and the line-of-sight velocity distribution (LOSVD).  This
routine works in pixel space and the LOSVD is determined by $\chi^{2}$
minimization.  The LOSVD is assumed to be in the form of a
Gauss-Hermite series, where the coefficients are determined 
simultaneously, but the solution is biased towards a Gaussian.

\section{Discussion}
\label{sec:disc}

\begin{figure}
\figurenum{1}
\plotone{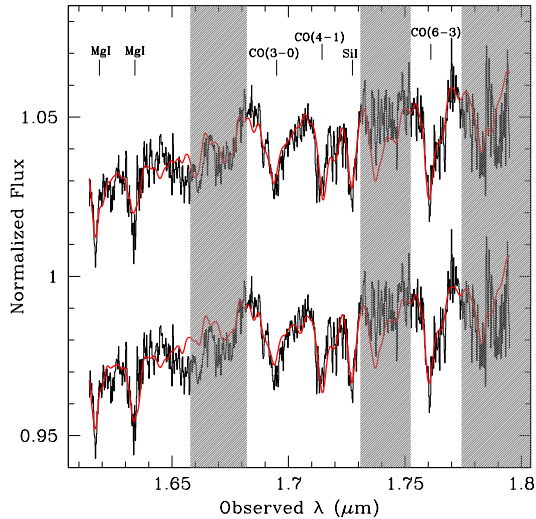}
\label{fig:spectra}
\caption{Two copies of the observed frame host-galaxy
  spectrum of PG1426+015, normalized and offset for clarity.  The
  smooth curve shows either the M5Ia (top spectrum) or K5III
  (bottom spectrum) velocity template, broadened to fit the
  host-galaxy absorption features using the pPXF method.  The gray
  bands show regions of the spectrum that were excluded from the fit
  for various reasons (see text).}
\end{figure}

Two copies of the host-galaxy spectrum of PG1426+015 are shown in the
top and bottom panels of Figure~\ref{fig:spectra}.  In the top
section, the smooth curve shows the spectrum of the M5Ia velocity
template, broadened by the best-fit LOSVD.  The bottom spectrum shows
the host spectrum with the broadened K5III velocity template.  No
systematic velocity offsets were present between the atomic and
molecular features in the host spectrum, so we fit all six absorption
features simultaneously.  The shaded areas designate spectral regions
excluded from our pPXF fit.  We excluded the $1.658-1.682 \mu {\rm m}$
region because it did not show absorption features and had moderate
telluric contamination, the $1.731-1.752 \mu {\rm m}$ region because
it was affected by dense sky emission, and the region redward of
$1.774 \mu {\rm m}$ because of increasingly strong telluric absorption
features.

The host-galaxy spectrum has an average signal-to-noise ratio (SNR)
of about 190 per pixel. This is larger than the SNR of the long-slit
ISAAC spectrum, which was obtained with five hours on-source
integration.  This increase in the SNR for a spectrum acquired in less
time is evidence of the combined advantages of the NIFS IFU and the
LGS AO system.

The velocity dispersion resulting from the M5Ia fit is $224 \, {\rm
km \, s^{-1}}$, with a reduced $\chi^{2}$ of 0.91, while the velocity
dispersion resulting from the K5III fit is $209 \, {\rm km \, s^{-1}}$,
with a reduced $\chi^{2}$ of 0.92.  The CO(3-0) feature at $1.693 \mu
{\rm m}$ is better fit by the M5Ia template while the \ion{Mg}{1}
feature at $1.632 \mu {\rm m}$ is better fit by the K5III template.
The M1III fit resulted in a velocity dispersion of  $207 \, {\rm km \,
s^{-1}}$, with a reduced $\chi^{2}$ of 0.98.  Because the M1III fit
was somewhat worse than the M5Ia and K5III fits, we do not include it
in Figure~\ref{fig:spectra} and we use the average of only the K5III
and M5Ia values as our final velocity dispersion.  The K0III was not a
good match to the host spectrum.  \citet{genzel01} found that K5-M1
giants and supergiants are the best templates for the young stellar
populations in ULIRGs \citep[also see][]{dasyra06a,dasyra06b} while 
D07 used K0-M0 giants and supergiants as templates for the presumably
older stellar populations in quasars. The fact that later
spectral-type templates provide better fits to the host spectrum of
PG1426+015 may point to a young stellar population in the host galaxy.
To determine the error in individual template fits, we varied the
velocity dispersion from its best-fit value while holding the velocity
and higher-order Gauss-Hermite coefficients constant.  In all three
cases (using the K5III, M1III, or M5Ia), the location of $\Delta
\chi^{2} = 1$ gave a $1\sigma$ error bar of $\sim 10~{\rm km \,
  s^{-1}}$.  As a measure of the error due to template mismatch, we
use the standard deviation of the velocity dispersions determined from
the three template fits, which is also $\sim 10~{\rm km \,
  s^{-1}}$. Our final error bar is the quadrature sum of these two
errors, resulting in a velocity dispersion for PG1426+015 of $217\pm15
\, {\rm km \, s^{-1}}$.

This measurement is within an effective physical aperture of 1.6~kpc.
For comparison, the effective radius ($r_{\rm e}$) for PG1426+015 is
4.86~kpc (based on a Galfit [Peng et al. 2002] analysis of Hubble
Space Telescope Wide-Field Planetary Camera 2 data).  Our aperture is
intermediate between $r_{\rm e}$ and $r_{\rm e}/8$, two commonly
quoted aperture sizes in the literature.  We consequently choose 
to use our measured value for the remainder of the analysis but note
that the velocity dispersion would be $208\, {\rm km \, s^{-1}}$ if we
corrected our value using the relation derived by \citet{jorgensen95}
for E and S0 galaxies.  This correction does not account for the 
contribution to the velocity dispersion that we miss by excluding the
central 0.16~kpc; however, this correction is expected to be small.
For comparison, D07 measured the velocity dispersion of PG1426+015 to
be $185\pm67\, {\rm km \, s^{-1}}$.  This is about $0.5\sigma$ lower
than our measured value, although these measurements are consistent.
Our higher precision is due to the higher SNR and broader wavelength
coverage of the NIFS data.

\begin{figure}
\figurenum{2}
\plotone{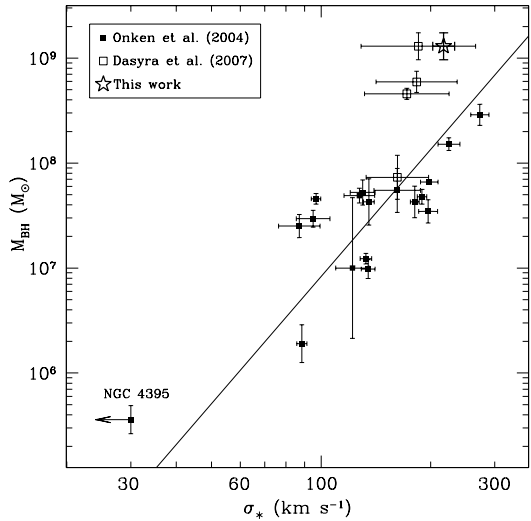}
\label{fig:m_sigma}
\caption{
The \msig\ relationship for reverberation-mapped AGNs. Filled squares
represent AGNs with velocity dispersion measurements based on the
\ion{Ca}{2} triplet.  These are from O04, updated with additional
velocity dispersion data and improved reverberation-based black hole
masses. Open squares show the PG quasars from D07, with velocity
dispersions measured from the CO bandheads. The open star represents
our new measurement, while the open square to its immediate left is
the D07 position of this object. The \citet{tremaine02} fit to the
quiescent galaxy \msig\ relationship is shown as a solid line.  We
assume $\langle f \rangle = 5.5$. 
}
\end{figure}

Figure~\ref{fig:m_sigma} shows that PG1426+015 is significantly above
the \msig\ relation for AGNs with masses determined by reverberation
mapping.  The filled squares represent \ion{Ca}{2} triplet
measurements from O04.  They have been updated with velocity
dispersion data from \citet{nelson04} and improved reverberation-based
black hole masses from \citet{bentz06,bentz07} and \citet{denney06}.
The open squares represent H-band CO bandhead measurements of the PG
quasars studied in D07.  The solid line denotes the \citet{tremaine02}
fit to the quiescent galaxy \msig\ relation.  We have not used the
\citet{ferrarese05} fit to the quiescent galaxy \msig\ relation
because $\langle f \rangle$ for the AGN population has not
yet been computed relative to this fit.  The velocity dispersion 
presented in this work moves PG1426+015 to the position marked by the
open star.  With our smaller error bar, PG1426+015 is now more
significantly discrepant with the \msig\ relation.  Note that if we
compare the location of PG1426+015 relative to the steeper
\citet{ferrarese05} \msig\ relation fit (and assume that $\langle f
\rangle$ is not significantly different from 5.5), it is somewhat less
significantly discrepant.

D07 found that three of their four reverberation mapped
PG quasars lie above the \msig\ relation and list three possible
reasons for this.  First, the scale factor, $f$, from
Equation~\ref{eqn:virial} could be different for PG quasars and the
lower-luminosity AGNs in the O04 sample.  In general, differences in
scale factors could arise if different populations have different
inclinations \citep[e.g.,][]{wu01,jarvis06}, accretion rates
\citep[e.g.,][]{collin06}, radiation pressure \citep{marconi08}, or
some combination of factors. Secondly, measurement errors in either
\sigs\ or $M_{\rm BH}$ could spuriously drive these objects to lie
above the \msig\ relation.  D07 note that underestimates
of the velocity dispersion by 10\% are expected because of quasar
continuum dilution.  We found that continuum dilution does not
significantly affect our measurement by varying our inner extraction
aperture to include more or less quasar continuum.  The velocity
dispersions resulting from these tests were not significantly
different from the value derived using the optimal inner aperture.  In
addition, preliminary results show that the black hole mass in
PG2130+099---one of the D07 quasars that lies above the \msig\ 
relation---may be overestimated by a factor of several, based on a new
measurement of the reverberation time lag (Grier et al., in
preparation). Thirdly, D07 show that small number statistics 
can at least partly account for the PG quasars' location above the
\msig\ relation.

Two final caveats should be considered.  First, this quasar is an
interacting system, with a companion at a nuclear separation of
$2.7\arcsec$ (4.4~kpc).  We would therefore expect the velocity
dispersion to be larger than predicted by the \msig\ relation, or
opposite to its observed location.  Additionally, CO and \ion{Ca}{2}
triplet velocity dispersions could give inconsistent results (see
Dasyra et al. 2006b for a discussion of this in the context of ULIRGs
and Silge \& Gebhardt 2003 for a discussion in the context of
early-type quiescent galaxies).  If PG1426+015 has an atypically young
stellar population in its nuclear region, as circumstantially
suggested by the need for later-type spectral templates, these young
stars may not trace the full velocity dispersion profile of the bulge
and cause an underestimate of the true stellar velocity dispersion.

\section{Summary}
\label{sec:conc}

We obtained H-band observations of PG1426+015 using the NIFS IFU on
Gemini North and the newly commissioned ALTAIR LGS AO system.  The
combination of using an IFU and LGS AO provides a unique tool for
studying the hosts of luminous quasars.  This is demonstrated by
the very high SNR spectrum of the quasar host galaxy presented in this
work as well as our precise measurement of the stellar velocity
dispersion.  We measure the stellar velocity dispersion of PG1426+015
to be  $217\pm15 \, {\rm km \, s^{-1}}$.

Our measured velocity dispersion and the \citet{peterson04} black hole
mass from reverberation mapping places this quasar significantly above
the \msig\ relation for local, quiescent galaxies. We have explored a
number of reasons why PG1426+015 might lie above the \msig\
relationship.  We can determine which of these reasons is correct by
measuring velocity dispersions for the remainder of the
reverberation-mapped PG quasars.  It would also be valuable to
reinvestigate the black hole masses of these objects and determine if
there are any systematic differences between  \ion{Ca}{2} and CO
velocity dispersion measurements. When these objectives have been
completed, we will be able to determine if these
higher-luminosity quasars lie on the \msig\ relation and if the
scale factor, $f$, is a function of luminosity.  Ultimately we hope to
include the higher-luminosity sample in the calculation of $\langle f
\rangle$ such that future black hole masses calculated using this
factor will be more reliable.

\acknowledgements
We thank Chad Trujillo, Tracy Beck, and Richard McDermid for
assistance with the Gemini observations and Doug Simons, the Gemini
Director, for granting us DD time to obtain additional template star
observations.  We also thank the anonymous referee for comments that
clarified this work. We acknowledge support for this work from NSF
through grant AST-0604066 to The Ohio State University.  L. C. W. is
supported by an NSF Graduate Fellowship.  Based on observations
obtained at the Gemini Observatory, which is operated by the
Association of Universities for Research in Astronomy, Inc., under a
cooperative agreement with the NSF on behalf of the Gemini
partnership: the National Science Foundation (United States), the
Science and Technology Facilities Council (United Kingdom), the
National Research Council (Canada), CONICYT (Chile), the Australian
Research Council (Australia), Minist\'{e}rio da Ci\^{e}ncia e
Tecnologia (Brazil) and SECYT (Argentina). These observations were
obtained for Program ID GN-2006B-SV-110.

\clearpage


\begin{thebibliography} {}

\bibitem[Bentz et al.(2006)]{bentz06} Bentz, M. C., et al. 2006, ApJ,
  651, 775

\bibitem[Bentz et al.(2007)]{bentz07} Bentz, M. C., et al. 2007, ApJ,
  662, 205

\bibitem[Blandford \& McKee(1982)]{blandford82} Blandford, R. D., \&
  McKee, C. F. 1982, ApJ, 255, 419

\bibitem[Boccas et al.(2006)]{boccas06} Boccas, M., et al. 2006, SPIE,
  6272, 114

\bibitem[Cappellari \& Emsellem(2004)]{cappellari04} Cappellari, M. \&
  Emsellem, E. 2004, PASP, 116, 138

\bibitem [Collin et al.(2006)]{collin06} Collin, S., Kawaguchi, T.,
  Peterson, B. M., \& Vestergaard, M. 2006, A\&A, 456, 75

\bibitem[Dasyra et al.(2006a)]{dasyra06a} Dasyra, K. M., et al. 2006a,
  ApJ, 638, 745

\bibitem[Dasyra et al.(2006b)]{dasyra06b} Dasyra, K. M., et al. 2006b,
  ApJ, 651, 835

\bibitem[Dasyra et al.(2007)]{dasyra07} Dasyra, K. M., et al. 2007,
  ApJ, 657, 102 (D07)

\bibitem[Denney et al.(2006)]{denney06} Denney, K. D., et al. 2006,
  ApJ, 653, 152

\bibitem[Elvis et al.(1994)]{elvis94} Elvis, M., et al. 1994, ApJS,
  95, 1

\bibitem [Ferrarese \& Ford(2005)]{ferrarese05} Ferrarese, L., \& Ford,
  H. 2005, Space Sci. Rev., 116, 523

\bibitem[Ferrarese \& Merritt(2000)]{fm00} Ferrarese, L. \&
  Merritt, D. 2000, ApJ, 539, L9

\bibitem[Ferrarese et al.(2001)]{f01} Ferrarese, L., Pogge,
  R. W., Peterson, B. M., Merritt, D., Wandel, A., \& Joseph,
  C. L. 2001, ApJ, 555, L79

\bibitem[Gebhardt  et al.(2000a)]{g00a} Gebhardt, K., et
  al. 2000a, ApJ, 539, L13

\bibitem [Gebhardt et al.(2000b)]{g00b} Gebhardt, K., et
  al. 2000b, ApJ, 543, L5

\bibitem [Genzel et al.(2001)]{genzel01} Genzel, R., Tacconi, L. J.,
  Rigopoulou, D., Lutz, D., \& Tecza, M. 2001, 563, 527

\bibitem[Graham et al.(2001)]{graham01} Graham, A. W., Erwin, P.,
Caon, N., \& Trujillo, I. 2001, \apjl, 563, L11

\bibitem[H${\rm \ddot{a}}$ring \& Rix(2004)]{haring04} H${\rm
  \ddot{a}}$ring, N., \& Rix, H.-W. 2004, ApJ, 604, L89

\bibitem[Herriot et al.(2000)]{herriot00} Herriot, G., et al. 2000,
  SPIE, 4007, 115

\bibitem[Jarvis \& McLure(2006)]{jarvis06} Jarvis, M. J., \& McLure,
  R. J. 2006, MNRAS, 369, 182

\bibitem[J${\rm \o}$rgensen et al.(1995)]{jorgensen95} J${\rm
  \o}$rgensen, I., Franx, M., \& Kj${\rm \ae}$rgaard, P. 1995, MNRAS,
  276, 1341 

\bibitem[Kormendy \& Richstone(1995)]{kormendy95} Kormendy, J., \&
Richstone, D. 1995, \araa, 33, 581

\bibitem[Marconi et al.(2008)]{marconi08} Marconi, A., Axon, D. J.,
  Maiolino, R., Nagao, T., Pastorini, G., Pietrini, P., Robinson, A.,
  \& Torricelli, G. 2008, ApJ, 678, 693

\bibitem[Marconi \& Hunt(2003)]{marconi03} Marconi, A., \& Hunt, L. K.
  2003, ApJ, 589, L21

\bibitem[McGill et al.(2008)]{mcgill08} McGill, K. L., Woo, J.-H.,
  Treu, T., \& Malkan, M. A. 2008, ApJ, 673, 703

\bibitem[McGregor et al.(2002)]{mcgregor02} McGregor, P., et al. 2002,
  SPIE, 4841, 178

\bibitem[Moorwood et al.(1998)]{moorwood98} Moorwood, A. F. M., et
  al. 1998, Messenger, 94, 7

\bibitem[Nelson et al.(2004)]{nelson04} Nelson, C. H., Green, R. F.,
  Bower, G., Gebhardt, K., \& Weistrop, D. 2004, ApJ, 615, 652

\bibitem[Onken et al.(2004)]{onken04} Onken, C. A., Ferrarese, L.,
  Merritt, D., Peterson, B. M., Pogge, R. W., Vestergaard, M., \&
  Wandel, A. 2004, ApJ, 615, 645 (O04)

\bibitem[Peng et al.(2002)]{peng02} Peng, C. Y., Ho, L. C., Impey,
  C. D., \& Rix, H.-W. 2002, ApJ, 124, 266

\bibitem[Peterson(1993)]{peterson93} Peterson, B. M. 1993, PASP,
  105, 247

\bibitem[Peterson et al.(2004)]{peterson04} Peterson, B. M., et al. 2004,
  ApJ, 613, 682

\bibitem[Silge \& Gebhardt(2003)]{silge03} Silge, J. D., \& Gebhardt,
  K. 2003, ApJ, 125, 2809

\bibitem[Tremaine et al.(2002)]{tremaine02} Tremaine, S., et al. 2002,
  ApJ, 574, 740

\bibitem[Vacca et al.(2003)]{vacca03} Vacca, W. D., Cushing, M.C, \&
  Rayner, J. T. 2003, PASP, 115,389

\bibitem[Wright et al.(1994)]{wright94} Wright, E. L., Eisenhardt, P.,
  \& Fazio, G. 1994, BAAS, 184(2503)

\bibitem[Wu \& Han(2001)]{wu01} Wu, X.-B., \& Han, J. L. 2001, ApJ, 561,
  L59

\end{thebibliography}
\end{document}